\documentclass[journal]{IEEEtran}
\usepackage{cite}

\usepackage{float}

\ifCLASSINFOpdf
  \usepackage[pdftex]{graphicx}
\else
\fi
\usepackage{amsmath}
\usepackage{algorithm}
\usepackage{algpseudocode}
\usepackage{caption}
\usepackage{subcaption}
\usepackage{array}
\usepackage{url}

\usepackage{booktabs}

\begin{document}

\title{Stencil Computations\\on Cerebras Wafer-Scale Engine}
\author{Elia~Belli and~Daniele~De~Sensi%
\thanks{Sapienza University of Rome, Italy}}

\markboth{Stencil Computations on Cerebras Wafer-Scale Engine}{}

\maketitle

\raggedbottom

\begin{abstract}
    Stencil computations are a fundamental kernel in scientific computing, critical for simulations in domains such as fluid dynamics and climate modeling. However, these computations are often memory-bound on traditional High-Performance Computing architectures like GPUs, struggling against the "Memory Wall". Simultaneously, the rise of AI-oriented hardware, such as the Cerebras Wafer-Scale Engine, offers massive core parallelism and high-bandwidth on-chip memory, though typically optimized for lower-precision workloads. This work investigates the viability of bridging this divergence by mapping stencil algorithms onto the Cerebras WSE-3.
    The study introduces CStencil, a novel framework designed to implement two-dimensional stencil computations on the WSE-3. To ensure a rigorous and fair performance evaluation, the research also adapts ConvStencil, a state-of-the-art GPU stencil solver, porting it from its original double-precision design to single-precision for execution on an NVIDIA A100 GPU.
    Experimental results show that the WSE-3’s distributed SRAM and mesh interconnect effectively eliminate the off-chip memory bottlenecks common in GPU implementations. CStencil achieves speedups of up to 342× over the adapted ConvStencil version. A roofline model analysis further confirms that CStencil saturates the available compute and memory resources, demonstrating that the WSE dataflow architecture can be successfully repurposed for traditional scientific algorithms. These findings highlight the potential of the WSE-3 to deliver hardware utilization levels unattainable on conventional systems, offering a promising path toward overcoming the memory limitations of current HPC architectures.
\end{abstract}


\section{Introduction}
\label{sec:introduction}

Graphics Processing Units (GPUs) have been the dominant architecture in High-Performance Computing (HPC) for the last decade, particularly for General-Purpose GPU (GPGPU) workloads.
Concurrently, the rapid emergence of Artificial Intelligence (AI) has driven an exponential growth in computational demand, doubling every 3-4 months, a pace that far exceeds Moore's Law.
This growing demand has necessitated a shift away from solely scaling general-purpose hardware toward domain-specific accelerators (DSAs). The primary operations dominating AI, such as matrix multiplications (GEMM, SPGEMM), are now being accelerated by integrating specialized units, like NVIDIA’s Tensor Cores Units (TCUs) and AMD’s Matrix Cores, into established GPU architectures.

A more radical strategy involves designing new hardware from the ground up, exemplified by architectures like Tenstorrent~\cite{tenstorrent}, Cerebras~\cite{CS3-datasheet}, SambaNova~\cite{sambanova}, TPUs~\cite{TPU} and GraphCore~\cite{graphcore}. For example, the Cerebras Wafer-Scale Engine (WSE), built on wafer-scale integration, features a massive array of specialized cores and a unique distributed memory model with unparalleled bandwidth, specifically designed for Machine Learning (ML) workloads.
This trend toward highly specialized hardware has created a potential divergence: while AI achieves record performance, other computationally intensive fields, such as traditional scientific computing, have not been the primary target of these new architectures. However, these paths can be re-converged. Many core scientific algorithms, such as stencil computations fundamental to physics simulations, are operationally analogous to the kernels for which these AI accelerators are optimized.

The central premise of this paper is that the performance benefits of mapping algorithms like stencils to AI-specific hardware can unlock new levels of efficiency by exploiting the accelerator’s high-bandwidth memory architecture. This strategy directly addresses the "Memory Wall" that constrains performance on conventional systems.
This work investigates this hypothesis by providing a detailed mapping of the 2D Stencil algorithm onto the Cerebras WSE-3 architecture, presenting CStencil as a novel implementation. We conduct a rigorous performance analysis against a state-of-the-art GPU implementation, ConvStencil \cite{convstencil}, including a comprehensive roofline analysis to isolate the performance-limiting bottlenecks across both architectures. Our results demonstrate that the memory architecture of the WSE-3 provides a significant advantage for memory-bound scientific workloads, pointing towards a re-convergence of specialized AI hardware and traditional HPC workloads.
\section{Background and State-of-the-art}
\label{sec:background}

\subsection{Stencil in Scientific Simulations}

Stencil computation is one of the seven critical numerical methods for scientific computing \cite{Berkeley}, foundational to solving problems in heat diffusion \cite{heat-equation}, wave propagation \cite{wave-equation}, and fluid dynamics \cite{cfd}. These phenomena are modeled by Partial Differential Equations (PDEs), which often lack analytical solutions (e.g., Navier-Stokes equations \cite{millennium-problems}) and are impractical or too expensive to study purely through experimentation (e.g., wind tunnels).
Scientific simulations, such as Computational Fluid Dynamics (CFD), address this by numerically solving PDEs through discretization. This process approximates the continuous domain with a discrete grid, transforming the PDE into an iterative system of equations solved on that grid. This iterative update is a stencil computation.

A critical consideration in these simulations is numerical precision. Lower-precision formats (e.g., FP16 and BF16) common in AI are faster but can introduce accumulating rounding errors that prevent convergence or lead to numerical instability. Consequently, many sensitive scientific simulations require at least single-precision (FP32) for stability, and some non-negotiably require double-precision (FP64). Since the WSE supports up to single-precision (FP32), it can be a viable architecture for many of these stability-sensitive applications.

\subsection{Stencil Computation}

A stencil computation iteratively updates the elements of a domain by applying a predefined pattern, or stencil kernel. At each iteration, a point is computed as a weighted function of its neighboring points.

Stencils are characterized by three aspects:
\begin{itemize}
    \item \textbf{Dimensionality}: The spatial dimension of the domain (1D, 2D, or 3D).
    \item \textbf{Shape}: Broadly categorized as Star Stencil (elements along axes) or Box Stencil (elements in a square/cube around a central point).
    \item \textbf{Radius}: The spatial range of considered elements.
\end{itemize}
 
The variety in stencil kernels, driven by the physics of the application, often necessitates specialized algorithms and automated code generators, such as DRStencil~\cite{DRStencil}, StencilPy~\cite{StencilPy}, GT4Py~\cite{GT4Py} and Devito~\cite{devito}. We focus here on manually written 2-dimensional stencils, which are the most commonly studied and the dimensionality targeted by our implementation, CStencil.

\begin{figure}[h]
    \centering
    \includegraphics[width=0.9\linewidth]{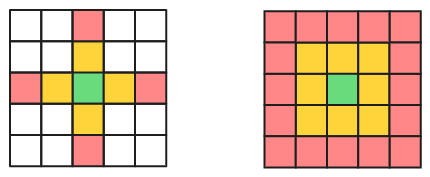}
    \caption{Examples of stencil kernels: Star2d-2r on the left and Box2d-2r on the right}
    \label{fig:stencil-shapes}
\end{figure}

\subsection{Jacobi Method}

The Jacobi Method is the most common iterative algorithm for solving stencil systems. The update rule at each point uses values exclusively from the previous iteration, making
the algorithm highly parallelizable. For an $m\times n$ grid, the update rule for a generic 2D stencil at interior point $u_{i,j}$ is:

\begin{align*}\label{jacobi-eq}
     u_{i,j}^{(k+1)} = \sum_{n\in \mathcal{N}} w_n\cdot u_{n_x,n_y}^{(k)}
 \end{align*}
where $k,k+1$ are the previous and current iterations, $\mathcal{N}$ represents the set of neighbors within the stencil footprint and $w_n$ is the specific weight associated with the neighbor at relative coordinates $(n_x,n_y)$.

In parallel implementations, the domain is decomposed into tiles distributed across cores. Adjacent tiles must exchange boundary values, a communication pattern known as \textit{halo swap}, though in practice this overhead is often limited, as only the boundary layer of each tile needs to be communicated. The deeper challenge is computational: stencil kernels typically have a low byte-to-flop ratio, making them memory-bound on classic multicore architectures. The Cerebras WSE's massive on-chip fabric addresses both concerns, providing high-bandwidth memory access alongside efficient on-chip communication.

\subsection{Related Work: GPU Stencil Implementations}

A current trend in HPC is the use of Tensor Core Units (TCUs) on GPUs to accelerate non-GEMM operations, including stencils.
The stencil can be mapped onto a GEMM by unrolling the kernel into a vector, analogous to the im2col transformation~\cite{im2row}, transforming the domain update into a matrix multiplication. The main issue with this approach is the high memory overhead arising from data redundancy, many domain points share neighbor data, leading to a highly sparse weight matrix. Despite the high TCU compute throughput, severe memory-bound nature limits their effectiveness.

State-of-the-art work has focused on optimization patterns to reduce this redundant memory traffic and better leverage TCUs. 
This exploration began with TCStencil~\cite{tcstencil}, which initially treated stencil computations as reduction operations; however, this approach was limited by the low precision of FP16 and inefficient memory access patterns. Addressing these shortcomings, ConvStencil~\cite{convstencil} introduced a convolution-based algorithm utilizing FP64 Tensor Cores. By implementing the "stencil2row" transformation and Dual Tessellation, ConvStencil reduced memory occupancy while significantly boosting TCU utilization. Building on these foundations, more recent works have introduced additional optimizations, such as LoRAStencil~\cite{LoRAStencil}, which employs low-rank decomposition for symmetrical kernels, and both SPTCStencil~\cite{SPTCStencil} and SparStencil~\cite{SparStencil}, which leverage structured sparsity to better align stencil patterns with Sparse TCUs.
Due to its high performance and open-source code availability, we select ConvStencil as the representative state-of-the-art GPU baseline for our evaluation.

\subsection{Related Work on Cerebras WSE}

Research on stencil computation for Cerebras systems has primarily targeted the CS-1 and CS-2 generations. Early work on the CS-1 \cite{CS-BigCGstab} explored 3D stencil solvers, using a key mapping strategy where one spatial dimension is fully contained within a single core, and the remaining two dimensions are tiled across the 2D wafer fabric. Subsequent work on the CS-2 \cite{25-point-stencil} adapted this strategy for high-order seismic modeling, transitioning to full single precision and a direct FMAC-style update for a 25-point star stencil. This implementation demonstrated near-perfect weak scaling and shifted the kernel from being traditionally memory-bound to compute-bound on the wafer.
While the previous CS-2 work compared performance against traditional GPU implementations that do not leverage TCUs, we shift our focus to the current state-of-the-art. Since TCU-based implementations now define the performance ceiling for GPUs, we select one of these advanced frameworks as our primary baseline for evaluation. 
Parallel to these performance-oriented efforts, there has also been work on developing higher-level abstractions and domain-specific languages to improve programmability on wafer-scale systems, a goal that frameworks such as SPADA~\cite{SPADA} explicitly target.

Crucially, most existing implementations target inherently three-dimensional stencils. While they can technically evaluate 2D problems, the mapping strategy restricts the problem size per core, resulting in inefficient utilization of the wafer's resources. To address this gap and efficiently support natively 2D workloads, we introduce CStencil, a framework specifically designed for two-dimensional stencils on the CS-3 with an optimized mapping that fully exploits the device’s compute density and on-chip memory capacity.
\section{Cerebras Wafer Scale Engine 3}
\label{sec:architeture}

\begin{figure}[h]
    \centering
    \includegraphics[width=0.9\linewidth]{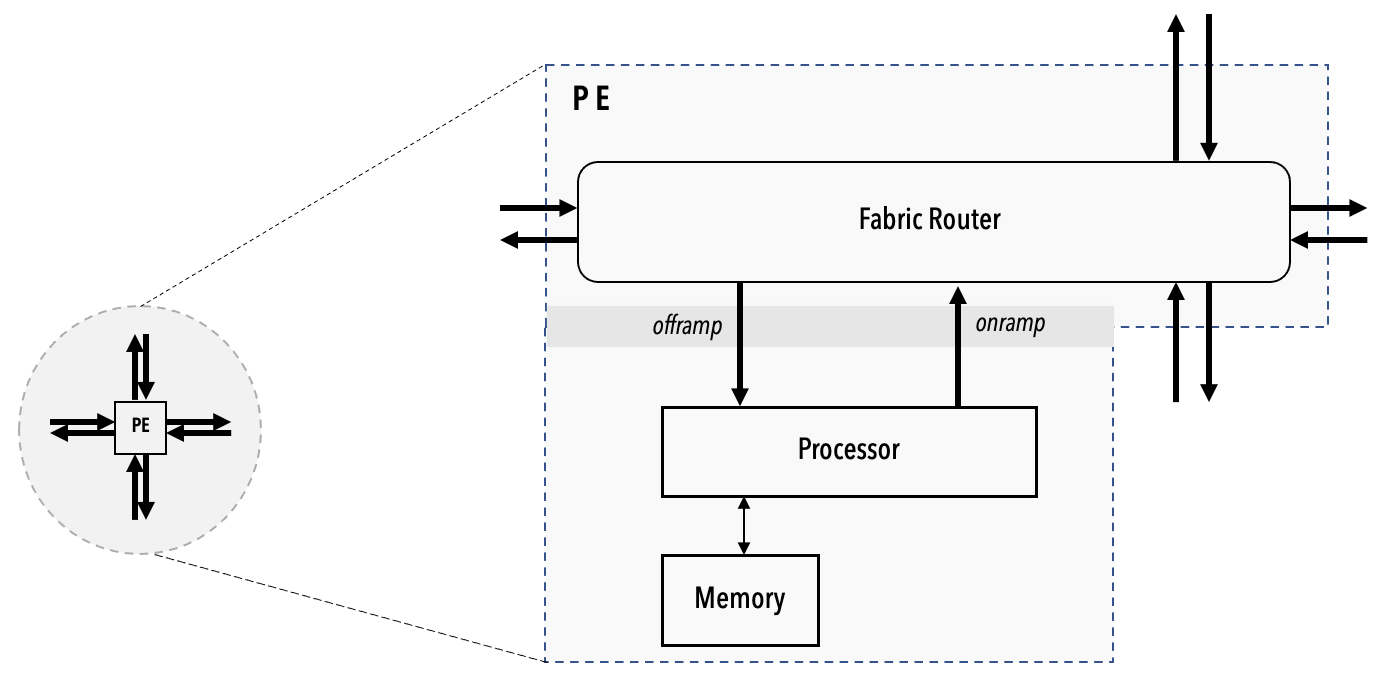}
    \caption{Components of a processing element (Source \cite{CS-SDK}).}
    \label{fig:pe}
\end{figure}

The Cerebras Wafer-Scale Engine 3 (WSE-3) represents a scale-up approach to computing, fabricating a massive parallel processor on a single silicon wafer. The WSE-3 is housed within the Cerebras System 3 (CS-3 \cite{CS3-datasheet}), which acts as the device and connects to a traditional CPU host via 100 Gb/s Ethernet connections. All interactions with the WSE, including data streaming and kernel launches, are orchestrated from the host via the Cerebras SDK \cite{CS-SDK}.
The WSE-3 is a fine-grained, data-flow architecture featuring 900,000 processing elements (PEs) interconnected in a two-dimensional rectangular mesh. Each PE can only communicate with its four immediate neighbors via 32-bit messages, called \textit{wavelets}, with an aggregate core-to-core bandwidth of 215 PB/s.
Each PE is independent and contains three main components (Fig. \ref{fig:pe}): a Compute Engine (CE), a Fabric Router, and dedicated SRAM.

\subsection{Distributed Memory}
Unlike traditional architectures that rely on slow, shared DRAM and cache hierarchies, the WSE-3 achieves its performance through a distributed memory model, placing SRAM directly next to the compute logic. The total on-chip memory is 44 GB, distributed as 48 KB of local SRAM per PE.
This design eliminates the need for cache management, allowing memory bandwidth to match the core data path. The 48 KB memory of a PE is laid out in 8 banks of 6 KB each, with each successive 32-bit word located in a successive bank. This layout allows
for a 192-bit access per cycle, with two 64-bit reads and one 64-bit write.

\subsection{Compute Engine and Execution Model}
The Compute Engine (CE) is the processor of the PE, and it can execute three types of operations: functions, tasks, and microthreads.
Functions are traditional procedures that execute immediately upon being called. Tasks, in contrast, cannot be called directly and must be activated instead.
Activation can occur by issuing a command directly, upon data reception, or when asynchronous operations complete. Activating a task places it in an activation queue, where it waits until the scheduler selects it for execution. Both functions and tasks are non-preemptive: once they start, they run to completion.
The CE also supports up to 8 active microthreads, although only one can execute at a time. Microthreads are always associated with asynchronous data transfer operations to or from the fabric. Once started, a microthread operation can be removed from execution. This occurs if the operation is reading from an empty fabric queue or writing to a full one, allowing other computation to proceed while the data transfer is blocked.
When the CE becomes idle (because a task has finished or a microthread has stalled), the hardware scheduler selects the next unit of work to execute by picking an eligible task from the activation queue or an active microthread.

The key abstraction for high-throughput computation on the CE is the Data Structure Descriptor (DSD). A DSD is a compact hardware descriptor that defines a strided memory region or a stream of wavelets. Concretely, a DSD specifies a base address, element count, and stride pattern, describing how to traverse a (possibly non-contiguous) chunk of memory. When an operation is issued on a DSD, the hardware automatically walks this pattern element-by-element without requiring explicit loop instructions.
DSDs are operated on via SIMD (Single Instruction, Multiple Data) instructions issued through Data Structure Registers (DSRs). When bank conflicts are avoided, which holds for the regularly strided access patterns typical of stencil computations, the hardware processes multiple elements per cycle, effectively replacing an entire nested loop with a single instruction. This abstraction is central to CStencil's computation strategy, described in Section \ref{subsec:computation}, where shifted DSDs are used to implement neighbor access patterns across the stencil kernel without any data rearrangement.

The CE is optimized for reduced-precision types such as \texttt{fp16}, \texttt{bf16}, and \texttt{cb16}. Full single-precision \texttt{fp32} operations such as \texttt{fma} and \texttt{fmul} are not natively supported in SIMD mode, though mixed-precision alternatives are available. CStencil exclusively uses \texttt{fp32} to maximize numerical accuracy.

\subsection{Fabric Router and Communication}
The router is the fundamental component enabling inter-core communication, serving as the sole mechanism through which Processing Elements (PEs) exchange data via wavelets. Each router is linked to its four immediate neighbors in the cardinal directions through bidirectional physical links, as well as to its local Compute Engine (CE). This design ensures that data arriving from the fabric can be consumed immediately by the CE for processing, supporting the WSE's native streaming execution model.

Data movement is managed through the concept of colors, which act as virtual channels with pre-defined input and output directions. By configuring multiple colors, a PE can simultaneously manage several independent communication paths. When a wavelet is injected into the fabric, it is placed into an output queue associated with a specific color, which dictates its routing path. These routing configurations are typically determined and loaded at compile time.

To support more sophisticated dataflow, the WSE allows for runtime reconfiguration of routing tables. This is achieved through specifically encoded wavelets that carry control commands for the router.
For example, a router can be pre-loaded with several distinct configurations; upon receiving a wavelet with an \texttt{ADV} (advance) command, the router switches its active state to the next configuration in the sequence. Because colors are a finite hardware resource, this dynamic mechanism is essential for implementing complex, non-static communication patterns across the wafer, though it introduces a significant increase in program design and control logic complexity.

\subsection{Programming Model}
Developing for the WSE involves a host-device paradigm where the computational kernel is written in the Cerebras Software Language (CSL) and the host orchestration is implemented via the SDK Python API. The compilation process via the \texttt{cslc} compiler targets the specific wafer generation (WSE-3) and produces an executable that maps the high-level logic to the physical fabric.
A program is fundamentally composed of three parts: the layout, which specifies the grid of PEs and their virtual interconnections; the host program, which orchestrates kernel launches and host-device communication; and the device program residing on the PEs.

Similar to classic host-device systems, the overall program execution is usually composed of three high-level steps: i) Streaming data to the device; ii) Launching the kernel on the device; iii) Copying the result back to the host. Since streaming data onto the WSE can take significantly more time than the computation itself, it is not viable for the host and device to communicate back and forth frequently. If any type of coordination is needed between the cores, it should not be orchestrated by the host; instead, it's preferable to code it implicitly, with potentially complex communication patterns.

The core execution unit on a PE is the \textit{Task}. Tasks are non-callable procedures initiated by one of three mechanisms: an explicit \texttt{@activate} command, the arrival of wavelets, or the completion of asynchronous Data Structure Descriptor (DSD) operations. During compile-time \textit{binding}, each task is assigned a unique Task ID associated with a specific hardware resource or communication channel. A task's execution is governed by its state (blocked or unblocked), which can be manipulated via \texttt{@block} and \texttt{@unblock} commands. If an activation command is issued for a blocked task, it will not be scheduled for execution until it is unblocked.

Tasks are categorized into three primary types:
\begin{itemize}
    \item \textbf{Data Tasks}: Wavelet-triggered tasks central to the data-flow architecture, activated upon the arrival of specific data wavelets.
    \item \textbf{Local Tasks}: Activated exclusively by the PE on which they reside, triggered by explicit commands or the completion of asynchronous DSD operations.
    \item \textbf{Control Tasks}: Specialized wavelet-triggered tasks activated by a control wavelet containing a specific Task ID in its payload.
\end{itemize}

A common use case for control tasks is the \textit{sentinel}, which signals the end of a data transmission. This is particularly useful for bursts of unknown length or for optimizing known-length transmissions by eliminating the need for continuous wavelet counting. Furthermore, control wavelets enable dynamic routing control; they can encode actions that instruct routers to update their configurations at runtime, allowing the program data to influence the network fabric itself.

\section{CStencil Design}
\label{sec:methodology}
CStencil implements an iterative 2D stencil solver based on the Jacobi Method on the WSE-3, supporting both Star and Box stencil patterns of various orders.
To execute the stencil computation, input data is pre-processed on the host to map effectively onto the processing elements. Once data preparation is complete, both the data and stencil weights are streamed onto the WSE. 
Host-device communication is absent during stencil iterations, save for periodic convergence checks. As these checks are infrequent enough to be considered negligible, our primary focus remains on the continuous execution blocks that occur in between.

Each iteration proceeds through two distinct phases:
\begin{itemize}
    \item \textbf{Communication (Halo Swap)}: Cores exchange the boundary elements (halo data) required by their neighbours to update their respective grid partitions;
    \item \textbf{Computation}: The core performs the actual stencil update on its local grid partition.
\end{itemize}

\subsection{Data Preparation} 
\begin{figure} [h]
        \centering
        \includegraphics[width=.9\linewidth]{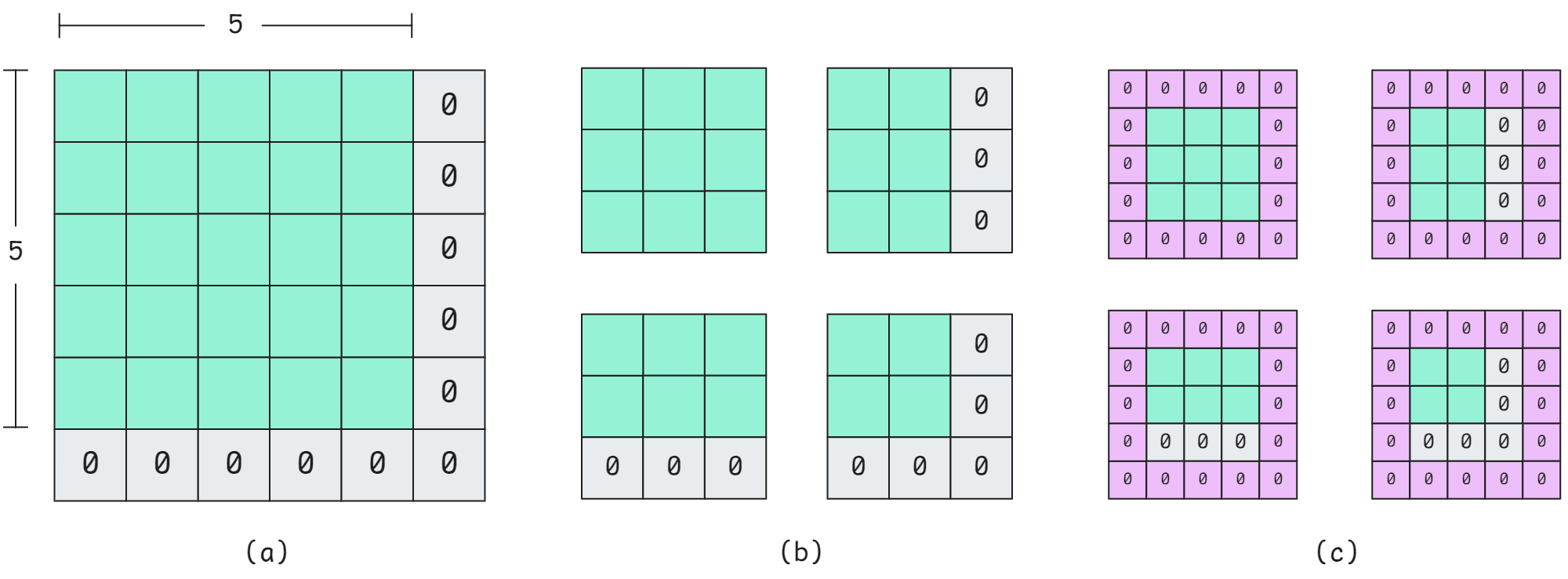}
        \caption{Visual representation of the padding and partitioning process for a $5\times 5$ input matrix distributed across a $2\times 2$ PE grid, for a stencil of radius $r=1$. The key features are: the original input grid (green cells), the global padding (grey cells), and the local halo padding (purple cells).}
        \label{fig:CStencil-mapping}
    \end{figure}

The data preparation phase is common to both stencil patterns and proceeds in three main steps:
    \begin{enumerate} 
        \item \textbf{Global Padding}: The global input matrix is initially padded with zeros, if necessary, to ensure its dimensions are evenly divisible by the PE grid dimensions (Fig.~\ref{fig:CStencil-mapping} (a)). This alignment guarantees that the matrix can be uniformly partitioned into equal sized sub-grids without altering the final output. Furthermore, it naturally enforces the zero boundary condition, as the PEs managing the global halo region maintain this zero padding throughout execution.
        \item \textbf{Grid Division}: The globally padded matrix is divided into an equal number of tiles (Fig.~\ref{fig:CStencil-mapping} (b)), such that each PE receives exactly one tile. 
        \item \textbf{Local Halo Padding}: The local tiles are subsequently padded with a zero halo (Fig.~\ref{fig:CStencil-mapping} (c)). The depth of this halo is determined by the stencil radius, $r$, and allocates the necessary memory to store neighbouring elements received during the halo exchange phase. Consequently, the halo thickness extends $r$ grid points on all four sides. This padding also serves as the zero boundary condition for PEs located at the edges of the global domain. 
    \end{enumerate}

\subsection{Communication Strategy}
CStencil's communication strategy relies on the constraint that the local sub-grid dimensions must exceed the stencil radius. This guarantees that all necessary halo elements reside exclusively on direct neighbours (including diagonals). This design choice simplifies the communication phase, eliminating the need to route data through intermediate PEs, thereby minimizing latency.
    
The communication mechanism is determined by the reach of the stencil, distinguishing between Star and Box patterns. Star patterns primarily require data exchange with neighbors in the cardinal directions (North, South, East, and West). In contrast, Box patterns extend this requirement to include diagonal points, necessitating data retrieval from halo corners situated on the diagonally adjacent PEs.
Since the Cerebras PEs are restricted to communicating directly only with their cardinal neighbours, the halo data for Box patterns must be forwarded to reach the diagonal neighbours.

\subsection{Star Pattern}
\begin{figure} [h]
    \centering
    \includegraphics[width=0.45\linewidth]{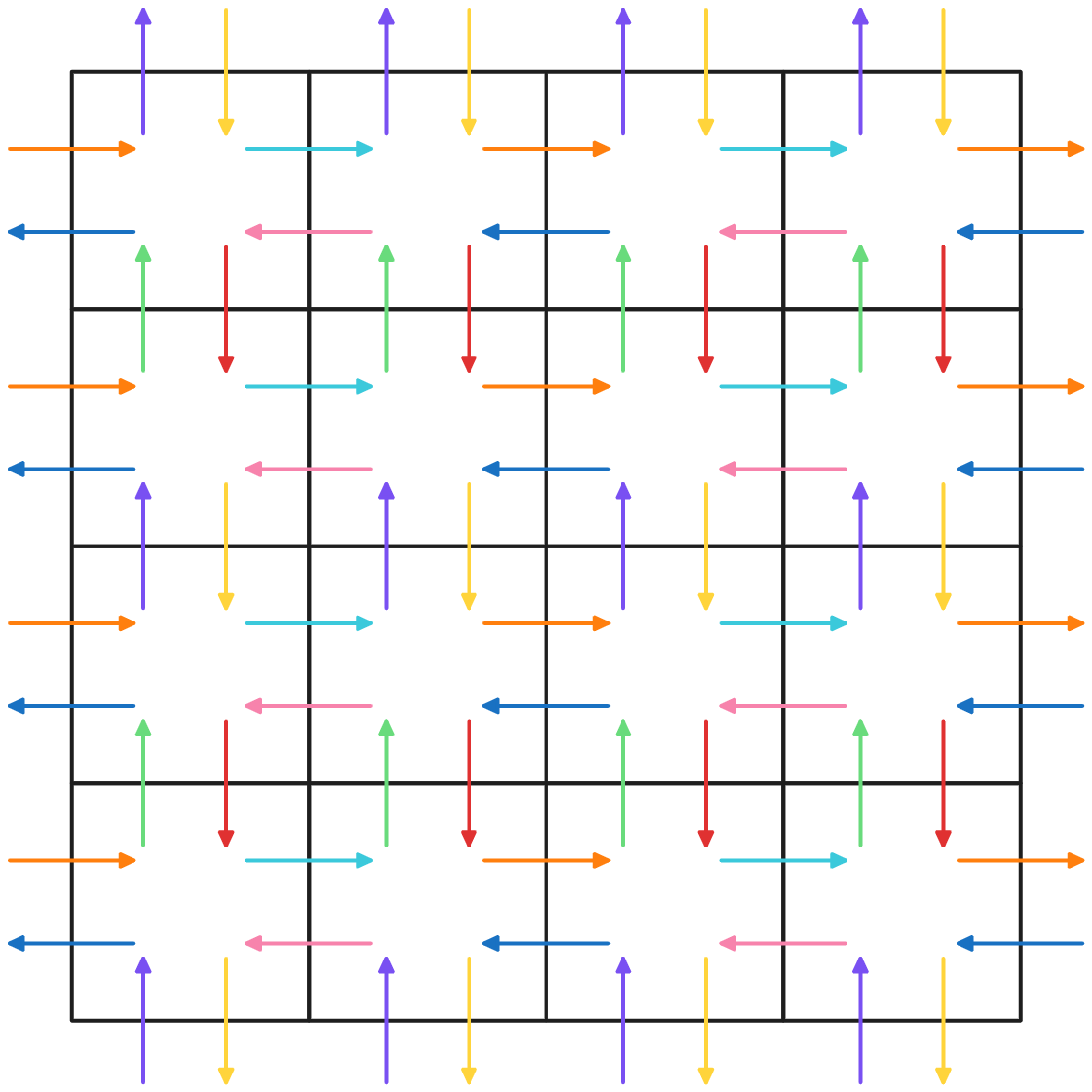}
    \caption{CStencil communication layout on a $4\times 4$ PE subarray. The configuration employs a checkerboard pattern to manage the 8 distinct communication colors (4 transmit, 4 receive) across the grid.}
    \label{fig:CStencil-layout}
\end{figure}

As established, the Star pattern relies exclusively on communication between direct cardinal neighbours. The communication scheme is symmetric: each PE sends its boundary data to its neighbours and receives the corresponding halo data in return. Consequently, the program layout necessitates $8$ distinct colors: $4$ for transmitting data in each cardinal direction and $4$ for receiving data from each direction, as shown in Figure~\ref{fig:CStencil-layout}.

\begin{figure} [h]
    \centering
    \includegraphics[width=0.45\linewidth]{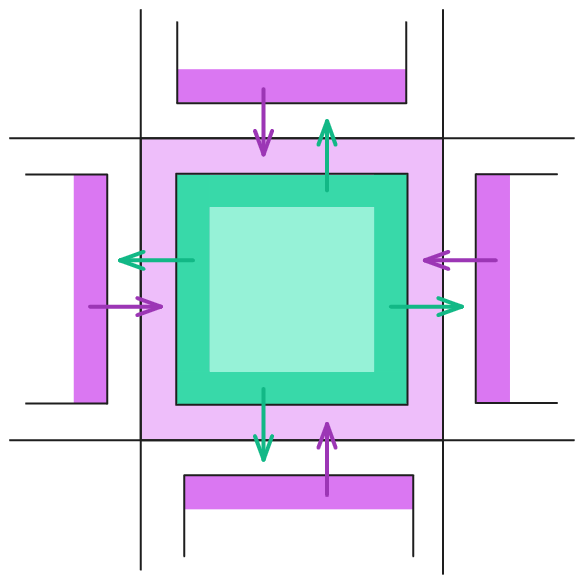}
    \caption{Visualization of the communication in a Star Pattern for one PE. The edges (green) are sent, the halo (purple) is received.}
    \label{fig:Star-communication}
\end{figure}

Therefore, the communication for a Star pattern consists of a single phase in which each PE exchanges boundary elements with its neighbours, as depicted in Figure~\ref{fig:Star-communication}. The implementation details of this communication protocol are described below.

\subsubsection{Asynchronous Transmission} 
To maximize bandwidth utilization, data transmission is implemented through asynchronous \texttt{@movs} operations managed by Data Structure Descriptors (DSDs). These DSDs are essential for handling grid data stored in row-major order with padding; their ability to define up to 4-dimensional access patterns allows Processing Elements (PEs) to transmit non-contiguous boundary edges in a single instruction, eliminating the need for costly data rearrangement. During execution, each asynchronous send occupies a hardware microthread; consequently, simultaneous communication in all four cardinal directions utilizes four of the six available microthreads, leaving the remaining two for system-level tasks.

\subsubsection{Data Reception and Synchronization} 
For data reception, a specific data task is configured for each cardinal direction. Upon the arrival of a data wavelet, the corresponding task writes the payload directly into the correct position within the local halo. Since the shape and volume of incoming data are static and known, the program could theoretically track completion by manually incrementing and checking a counter upon every wavelet arrival.
However, while a single counter check is computationally inexpensive, performing it for every individual wavelet introduces significant cumulative overhead, particularly for higher-order stencils involving large data volumes. This consumes instruction issue slots that could otherwise be utilized for compute.
Instead, we can make use of an asynchronous DSD operations functionality: an asynchronous operation can be set up to automatically activate a local task upon its completion.
This, coupled with the sentinel functionality of control tasks, allows us to notify the PE of send completions. Together, these mechanisms effectively implement a synchronization strategy to ensure that the halo swap phase is completed correctly.

The notification mechanism is coordinated through a series of hardware-triggered events: first, the completion of an asynchronous \texttt{@movs} send operation triggers a local task on the sending PE, which increments a local send counter and dispatches a control wavelet to the receiving neighbor. Upon arrival, this control wavelet activates a sentinel task that updates a local receive counter on the destination PE. By tracking these counters against the total number of active neighbors, the PE can verify that all data transfers are finished before transitioning from the synchronization phase to the computation phase.

\subsubsection{Necessity of Send-Completion Synchronization} 
Although a Processing Element could technically start computing once its halo is received, the non-preemptive nature of Cerebras tasks requires a strict synchronization barrier. If a PE begins a compute task while asynchronous sends are still in progress, the microthreads managing those transmissions are paused, forcing neighboring PEs to idle until the fast PE finishes its computation and resumes sending data. This creates a dependency chain that introduces execution bubbles equivalent to the duration of the compute phase. By enforcing a barrier where both senders and receivers must complete, CStencil prevents these stalls and ensures neighbors remain synchronized across iterations.

\subsection{Box Pattern}
Using only the cardinal communication mechanism employed for the Star pattern would result in missing the halo corner elements needed by the Box stencil kernel. These corner elements are owned by the PEs diagonally adjacent to the computing PE.
To send this diagonal data, the only viable option is to pass it through a PE that is a common cardinal neighbour of both the sending PE and the receiving PE. This PE will then forward the data.

\subsubsection{Forwarding Challenges}
Theoretically, the optimal method for handling this forwarding is to leverage the WSE routing capability to directly forward wavelets using a dedicated color. This approach avoids forcing data "down the ramp" into the PE's local memory and subsequently "back up" via a new send operation.
However, implementing a direct router-forwarding strategy atop the current communication mechanism presents several challenges:

\begin{itemize}
    \item \textbf{Bandwidth Waste}: Multicasting an entire edge via the router to capture a single corner is inefficient, as the receiving PE must filter out a large volume of unneeded data.

    \item \textbf{Redundant Sends}: Using a dedicated channel for the corner avoids bandwidth waste but results in transmitting the same data twice, once as part of the edge and once independently. Attempting to split the transmission into a "cornerless" edge and a separate corner is constrained by hardware resource limits.

    \item \textbf{Limited Microthreads}: The WSE-3 has only 6 available hardware microthreads. Since 4 are already occupied by cardinal exchanges, there is insufficient capacity to handle additional asynchronous forwarding tasks. This forces a shift to sequential execution, which removes the performance benefit of direct forwarding.
\end{itemize}

\subsubsection{2-Stage Forwarding}
    
Due to the aforementioned constraints, CStencil employs a simpler, more robust store-and-forward strategy. This approach reuses the existing send/receive infrastructure by dividing the communication into two distinct stages:

\begin{enumerate}
    \item \textbf{Side Exchange}: Each PE exchanges its halo edges (which include the corners) with its cardinal neighbouring PEs, identical to the standard Star pattern communication.
    \item \textbf{Corner Forwarding}: Once the corner data resides in the local memory of the cardinal neighbours (the intermediary PEs), a second, separate exchange is initiated to forward only these corner elements to the final diagonal receiving PE.
\end{enumerate}

Although this approach is theoretically suboptimal in terms of latency, requiring an extra synchronization step and involving redundant data movement, it is significantly simpler to implement and consumes fewer critical hardware resources. Furthermore, for low-order stencils, the corner data represents a negligible fraction of the total halo, introducing minimal additional delay.

\begin{figure}[h]
        \centering
        \includegraphics[width=.9\linewidth]{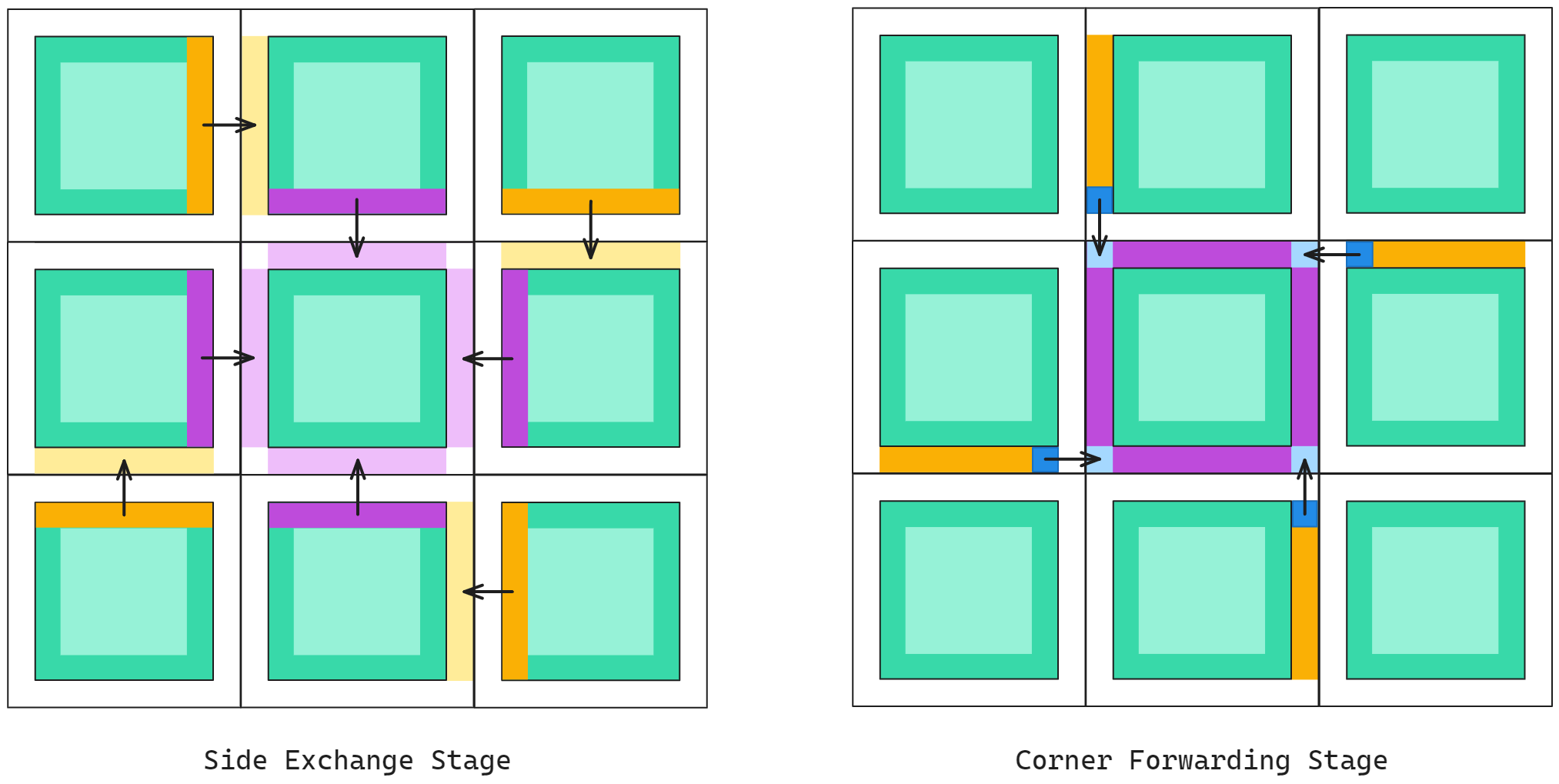}
        \caption{Visualization of the 2-stage communication in a Box Pattern. Only the data involved in the halo of the central PE is highlighted. In the Side Exchange Stage: purple edges represent standard communication as in the Star Pattern, while yellow edges contain the corners for the next stage. In the Corner Forwarding Stage: only the corners (blue squares) are forwarded to complete the communication.}
        \label{fig:Box-communication}
    \end{figure}

The first synchronization step mirrors the standard approach, with the primary difference being that a new communication phase is initiated once the barrier is reached. The second synchronization step is analogous, marking the completion of the forwarding phase and triggering the start of the computation.
In principle, the second stage could allow a single neighbour to transmit two corners to the receiving PE; for instance, the East PE could forward both the North-East and South-East corners. While simpler to implement, this approach fails to utilize the full bandwidth, leaving the North and South channels idle. Alternatively, a checkerboard pattern, where specific PEs use vertical channels to send and horizontal ones to receive (and vice versa), could be employed. However, this still leaves the network's full-duplex capabilities underutilized.
To fully exploit every channel in both directions simultaneously, we employ a rotational pattern, as illustrated in Figure \ref{fig:Box-communication}. This ensures that every neighbouring PE is actively involved in the corner forwarding stage. 

\subsection{Computation}
\label{subsec:computation}
    The core of the computation relies on the stencil kernel, which updates each grid point based on a weighted sum of its neighbours. In a standard scalar implementation, this requires iterating over the entire matrix with nested loops, explicitly fetching each neighbor, and updating points one by one. This approach introduces significant instruction overhead due to loop management and individual memory accesses.

    To overcome these bottlenecks, we exploit DSDs operations to describe the stencil computation not as a series of scalar accesses, but as element-wise matrix operations. This allows us to issue a single instruction that operates on the entire grid. The difference in complexity is illustrated in Listing \ref{lst:naive_vs_dsd}.

\begin{figure}[ht]
    \centering

    \begin{subfigure}[t]{0.48\textwidth}
        \hrule \vskip 5pt
        \begin{algorithmic}[1]
            \For{$i = 1$ to $H$}
                \For{$j = 1$ to $W$}
                    \State $acc \gets grid[i][j] \cdot w_c$
                    \State $acc \gets acc + grid[i-1][j] \cdot w_n$
                    \State $acc \gets acc + \dots$
                    \State $out[i][j] \gets acc$
                \EndFor
            \EndFor
        \end{algorithmic}
        \vskip 5pt \hrule
        \caption{Naive Scalar Implementation}
    \end{subfigure}
    \hfill
    \begin{subfigure}[t]{0.48\textwidth}
        \hrule \vskip 5pt
        \begin{algorithmic}[1]
            \State \textbf{Config} DSDs for Grid, Out
            \State 
            \State \texttt{@fmuls} $Out, Grid, w_c$
            \State \texttt{@fmacs} $Out, Grid_{north}, w_n$
            \State \texttt{@fmacs} $Out, Grid_{south}, w_s$
            \State \texttt{@fmacs} $Out, Grid_{east}, w_e$
            \State \texttt{@fmacs} $Out, Grid_{west}, w_w$
        \end{algorithmic}
        \vskip 5pt \hrule
        \caption{DSD Vectorized Implementation}
    \end{subfigure}

    \vspace{10pt}
    \caption{Comparison between naive scalar (left) and vectorized DSD (right).}
    \label{lst:naive_vs_dsd}
\end{figure}

\begin{figure} [h]
    \centering
    \includegraphics[width=1\linewidth]{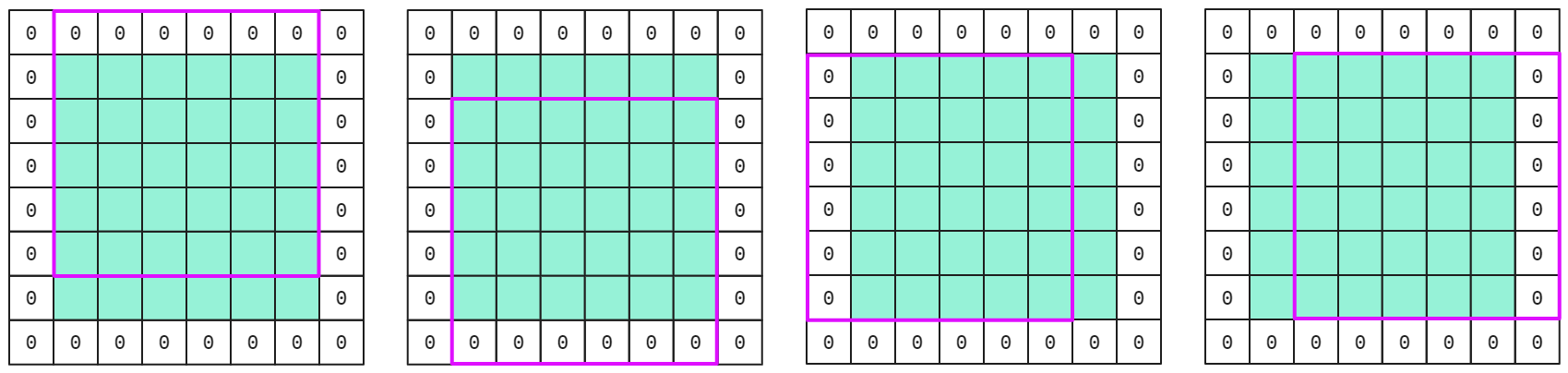}
    \caption{Visualization of the submatrices providing the North, South, West, and East contributions to the stencil (from left to right).}
    \label{fig:fmacs}
\end{figure}

The computation is decomposed into a sequence of element-wise matrix operations, each corresponding to a weight in the stencil kernel. As shown in Figure~\ref{fig:fmacs}, each neighbor submatrix is obtained by shifting the DSD's base pointer by one row or column, aligning the neighbor values with their respective center cells across the entire grid. The accumulator is first initialized by multiplying the center-aligned submatrix by $w_c$ via \texttt{@fmuls}. Each subsequent \texttt{@fmacs} adds a weighted neighbor contribution: for the North weight $w_n$, the DSD starts one row above, aligning $Grid_N$ with the center cells; the same shift principle applies for South, East, and West. Since each DSD operation walks the full grid in a single instruction, the entire stencil update reduces to the five operations shown in Listing~\ref{lst:naive_vs_dsd}.

\section{ConvStencil Single-Precision Porting}
Since the WSE lacks hardware support for double-precision arithmetic, we ported ConvStencil~\cite{convstencil}, originally designed for 64-bit operations, to single-precision to ensure a fair comparison. This adaptation optimizes the existing GPU kernel for TF32 Tensor Core Units (TCUs) on the NVIDIA A100 architecture, maintaining the original \textit{stencil2row} transformation pattern and optimizing the \textit{Dual Tessellation} algorithm.

The transition from double to single precision is primarily constrained by the structural requirements of the Tensor Core fragments. Since the NVIDIA A100 lacks native support for pure single-precision (FP32) Matrix Multiply-Accumulate (MMA) instructions, the implementation utilizes a mixed-precision approach: multiplicands are loaded in \texttt{tf32} format, while the accumulator and output fragments maintain standard \texttt{float} precision. Further details regarding fragment dimensions and MMA operations are provided in Appendix \ref{app:mma_operations}.

\subsection{Stencil2Row Layout}
\begin{figure} [h]
    \centering
    \includegraphics[width=0.9\linewidth]{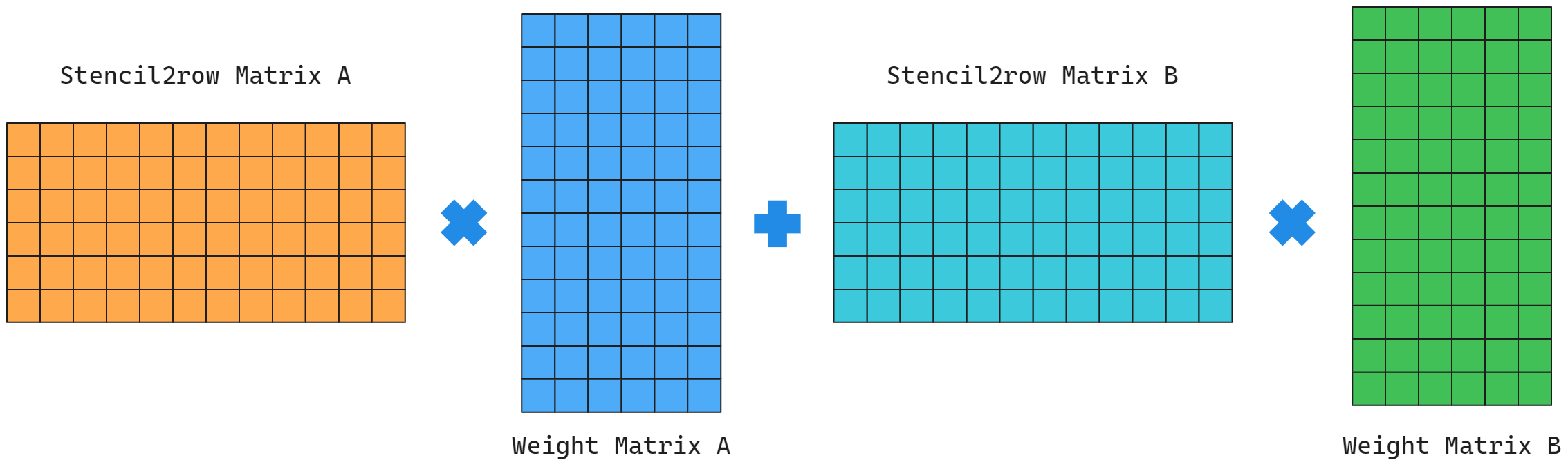}
    \caption{Stencil2row transformation results and visualization of the Dual Tessellation}
    \label{fig:conv-overview}
\end{figure}

ConvStencil applies the \textit{stencil2row} layout to an input grid, producing two distinct matrices, referred to as Stencil2row Matrix A and Matrix B, which encapsulate the data redundancies inherent to the stencil pattern. As illustrated in Figure~\ref{fig:conv-overview}, these matrices are multiplied by corresponding weight matrices, constructed using a complementary pattern, and subsequently accumulated to produce the final result.

\subsection{Dual Tessellation} 
ConvStencil core computation is the \textit{Dual Tessellation} algorithm, in the original double-precision kernel, a single warp computes one Dual Tessellation result by accumulating these two separate GEMMs results, referred to as \textit{Vitrolites}, performed as a sequence of $13$ MMA operations, into a single $\mathbf{C}$ fragment. Mathematically, a single warp computes the result $\mathbf{C}$ by sequentially accumulating these products:

\[
    \mathbf{C}_{8\times 8} = \underbrace{\sum_{k=0}^{12} \left( \mathbf{A}^{(A)}_k \mathbf{B}^{(A)}_k \right)}_{\text{Vitrolite A}}
    + \underbrace{\sum_{k=0}^{12} \left( \mathbf{A}^{(B)}_k \mathbf{B}^{(B)}_k \right)}_{\text{Vitrolite B}}
\]
    
where $\mathbf{A}^{(v)}_k$ and $\mathbf{B}^{(v)}_k$ represent the $k$-th tile loaded for Vitrolite $v \in \{A, B\}$. The computation is naturally sequential: the warp first computes the contribution of Vitrolite A and subsequently accumulates the contribution of Vitrolite B into the same accumulator.

\subsection{Packing Dual Tessellations}
In the single-precision port, the \texttt{tf32} fragments sizes are doubled compared to their \texttt{double} counterparts: $\mathbf{A}$ fragments are $16 \times 8$ and $\mathbf{B}$ fragments are $8 \times 16$. This increased capacity allows us to execute two Dual Tessellations together.
This packing strategy alters the kernel's work distribution by assigning a single warp the responsibility of computing two full Dual Tessellations ($DT_1$ and $DT_2$) simultaneously. 
Based on the Vitrolite formulation defined previously, this corresponds to four logical GEMMs computed in two execution phases, accumulated into the same $\mathbf{C}_{16\times 16}$ fragment.

We leverage the dimensions of \texttt{tf32} fragments to maximize throughput by packing original fragments along both the M-axis (batching independent tessellations) and the K-axis (aggregating original MMA operations).
\begin{itemize} 
    \item \textbf{Fragment A ($16\times8$):} Partitioned horizontally into $8\times8$ blocks, $DT_1$ (upper) and $DT_2$ (lower). Each block is formed by horizontally stacking two $8\times4$ tiles: $(A_k,A_{k+1})$ for $DT_1$ and $(A_k',A_{k+1}')$ for $DT_2$. 
    \item \textbf{Fragment B ($8\times16$):} Partitioned vertically into two $8\times8$ blocks. The left block is populated by vertically stacking two $4\times8$ weight tiles $(B_{k},B_{k+1})$, while the right half is filled with zero padding ($\mathbf{0}$). 
\end{itemize}

The packing along the K-axis allows the kernel to perform two accumulation steps per instruction. Consequently, the number of MMA iterations per Vitrolite is halved from $13$ to $7$ (loop $k=0, \dots, 6$).
Let $\mathbf{C}^{(k)}$ be the partial result from a single MMA step. The resulting block matrix multiplication produces a redundant output structure:
\[
\mathbf{C}^{(k)} = \mathbf{A}_{\text{packed}}^{(k)} \times \mathbf{B}_{\text{packed}}^{(k)}
=
\begin{bmatrix}
    \mathbf{C}^{(k)}_{DT_1} & \mathbf{0}\\
    \mathbf{C}^{(k)}_{DT_2} & \mathbf{0}
\end{bmatrix}
\]
    
\begin{figure}[h]
    \centering
    \includegraphics[width=0.9\linewidth]{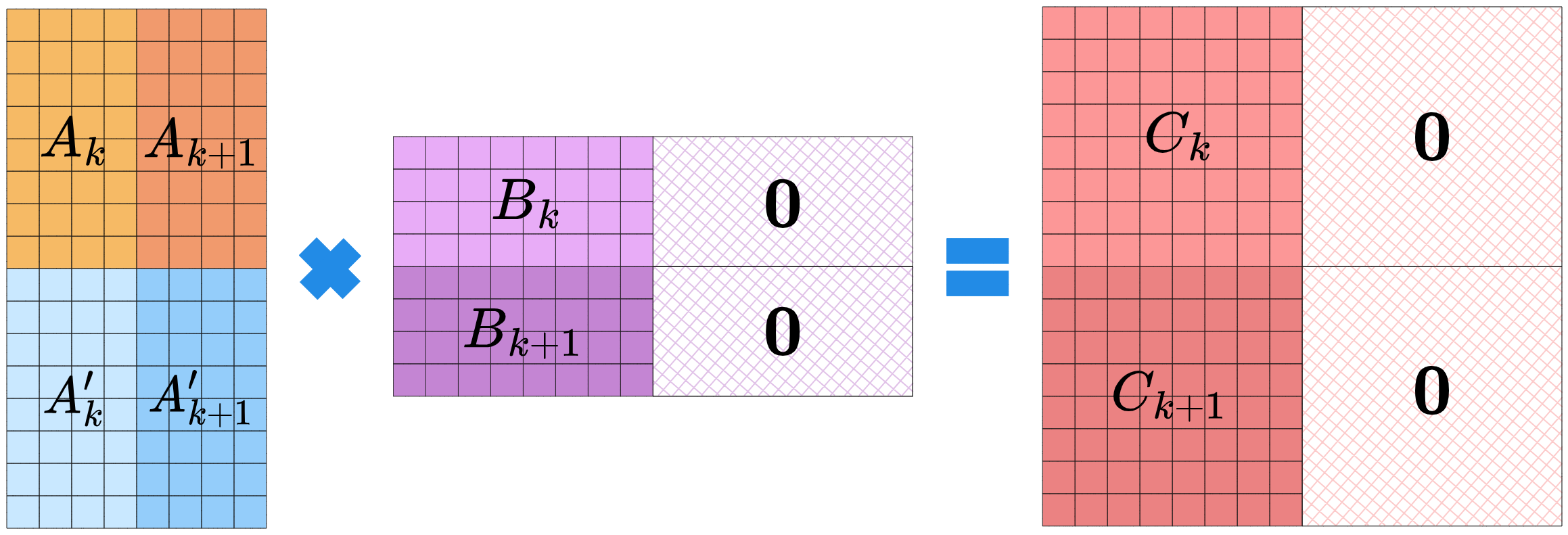}
    \caption{Packing strategy. The left operand packs four tiles: two from the current loop iterations ($k, k+1$) for $DT_1$ and two for $DT_2$. The right operand packs the corresponding weight tiles. The resulting accumulator isolates the computed values for two independent Dual Tessellations.}
    \label{fig:mma-packing}
\end{figure}

After summing the partial results from both execution phases (Vitrolite A and Vitrolite B), the final accumulator $\mathbf{C}_{16\times 16}$ contains the independent results for the two Dual Tessellations. This packing strategy is illustrated in Figure~\ref{fig:mma-packing}.
To finalize the operation, the accumulator $\mathbf{C}_{16\times 16}$ is stored in shared memory. Warp threads then collectively write the upper-left tile ($\mathbf{C}_{DT_1}$) and the lower-left tile ($\mathbf{C}_{DT_2}$), to global memory.

\subsection{Limitations}
The current packing strategy results in a 50\% underutilization of Tensor Core compute capacity due to a structural mismatch between the hardware dimensions and the data requirements. To satisfy the $8\times16$ requirement of Fragment B in \texttt{tf32} precision, the right half is zero-padded ($\mathbf{B}_{\text{packed}} = [\mathbf{B}_{\text{weights}} | \mathbf{0}]$), which forces the hardware to spend half of its floating-point operations calculating null values ($x\cdot 0$).

The final accumulator reflects this inefficiency, producing a valid left half and a zeroed right half: $\mathbf{C}_{16\times 16} = [\mathbf{C}_{\text{valid}} | \mathbf{0}]$. Unlike standard GEMM operations, the \textit{Dual Tessellation} algorithm pairs each input tile $A_i$ with a single specific weight tile $B_i$, lacking the independent weight sets needed to occupy the full width of the hardware fragment. Overcoming this gap requires redesigning the stencil-to-matrix mapping to expose further parallelism along the weight dimension.

\section{Evaluation}
\label{sec:evaluation}

\subsection{Experimental Setup}
To ensure a fair comparison, experiments were standardized across an NVIDIA A100 GPU (Ampere, 108 SMs, 64~GB HBM2e) and a Cerebras CS-3 (WSE-3). The GPU baseline utilized CUDA 12.2 with maximum optimization (\texttt{-O3}), while the wafer-scale implementation was developed using the Cerebras SDK 1.4.0.

\textbf{WSE Setup} Performance measurements for the WSE were obtained using the Cerebras cycle-accurate simulator. As discussed in other works~\cite{Luczynski_2024}, because threads can't be preempted, programs exhibit deterministic, state-machine-like behavior which can be modeled with a cycle-accurate simulator. 
Still, to ensure reliability of the results, the Star2d-1r kernel was executed on the actual CS-3 hardware; the resulting performance closely matched the simulator's predictions, validating its use for the remaining kernel variants.

\textbf{Benchmarking Methodology} The evaluation follows a weak-scaling study using a heat-diffusion simulation. The workload consists of 1,000 iterations on grids ranging from 128$\times$128 to 32,768$\times$32,768. This setup allows for a direct comparison of throughput and scaling efficiency between ConvStencil and CStencil.

\textbf{Metrics}
Performance is quantified using \textit{GStencil/s}, representing the number of grid cell updates per second. For a 2D grid with $T$ iterations and dimensions $N_x, N_y$, it is defined as:

    \begin{equation}
        \text{GStencil/s} = \frac{T \times N_x \times N_y}{t \times 10^9}
    \end{equation}

The execution time $t$ isolates the pure kernel runtime on both the WSE and the GPU, excluding initialization and host-device data transfer overheads. For the WSE-3, measurements are derived from hardware performance counters or cycle-accurate simulation results, converted to seconds using the nominal clock frequency of 875.0\,MHz.

\subsection{ConvStencil Port Comparison}

\begin{figure}[h]
        \centering
        \includegraphics[width=0.9\linewidth]{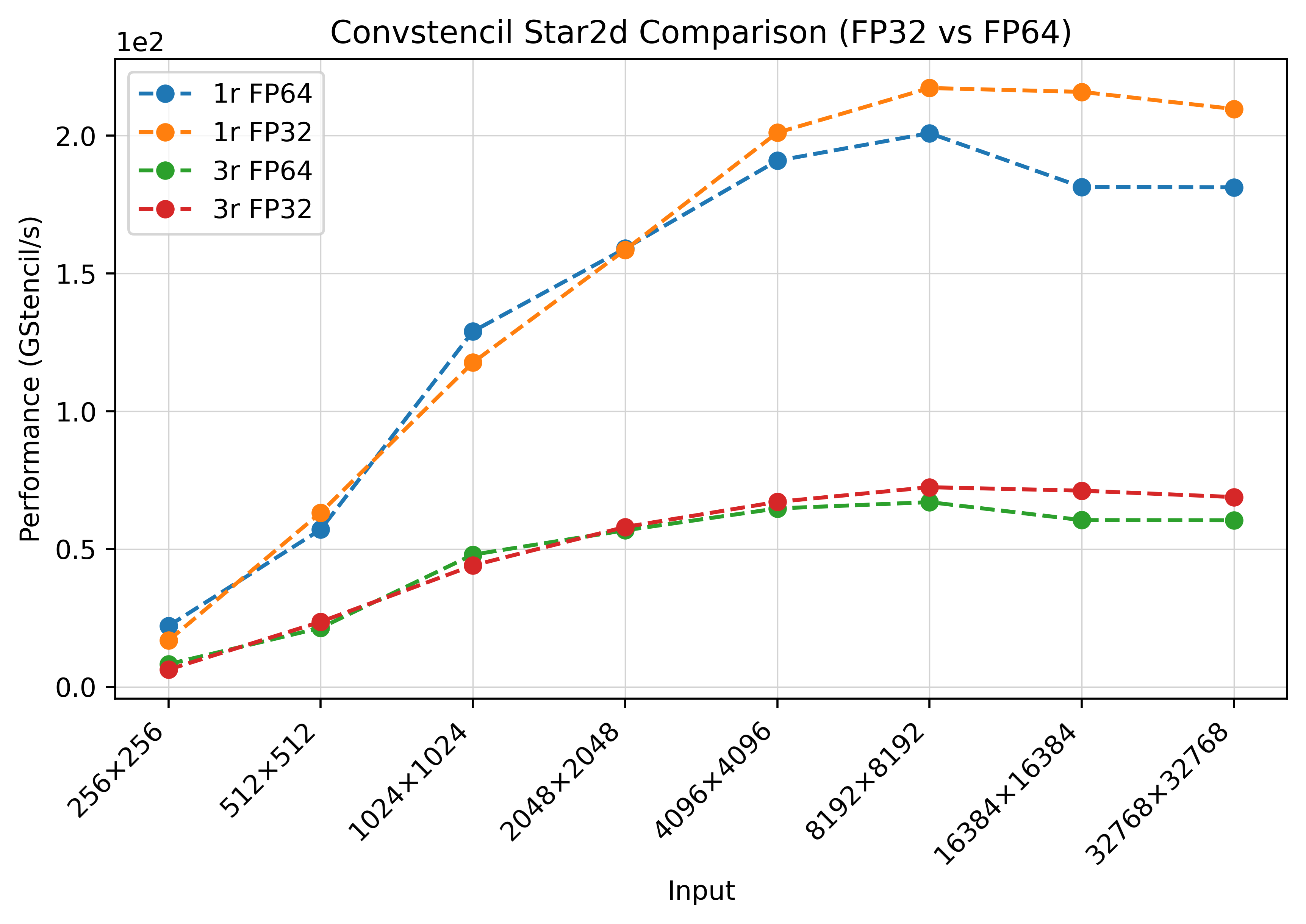}
        \caption{Weak scaling comparison between the original ConvStencil (Double Precision) and the single-precision port for both Star2d-1r and Star2d-3r patterns.}
        \label{fig:gpu_weak_scaling}
    \end{figure}

Figure~\ref{fig:gpu_weak_scaling} compares the weak scaling performance of the original double-precision ConvStencil and our single-precision port. Despite the theoretical 8$\times$ peak throughput advantage of TF32 Tensor Cores (156~TFLOP/s) over FP64 units (19.5~TFLOP/s), the single-precision port exhibits nearly identical performance to the original.

This parity is driven by two primary factors. First, the \textit{stencil2row} transformation results in an algorithmic limitation where approximately 50\% of the compute operations are multiplications by zero. Second, the kernel is strictly memory-bound; profiling shows that Tensor Core utilization dropped from 60\% in the double-precision version to 20\% in the single-precision port. Because TF32 units process data significantly faster, they exacerbate the existing data-supply bottleneck, as the memory subsystem cannot satisfy the increased demand for operands.

\subsection{CStencil Evaluation}
\begin{figure}[h]
        \centering
        \includegraphics[width=0.9\linewidth]{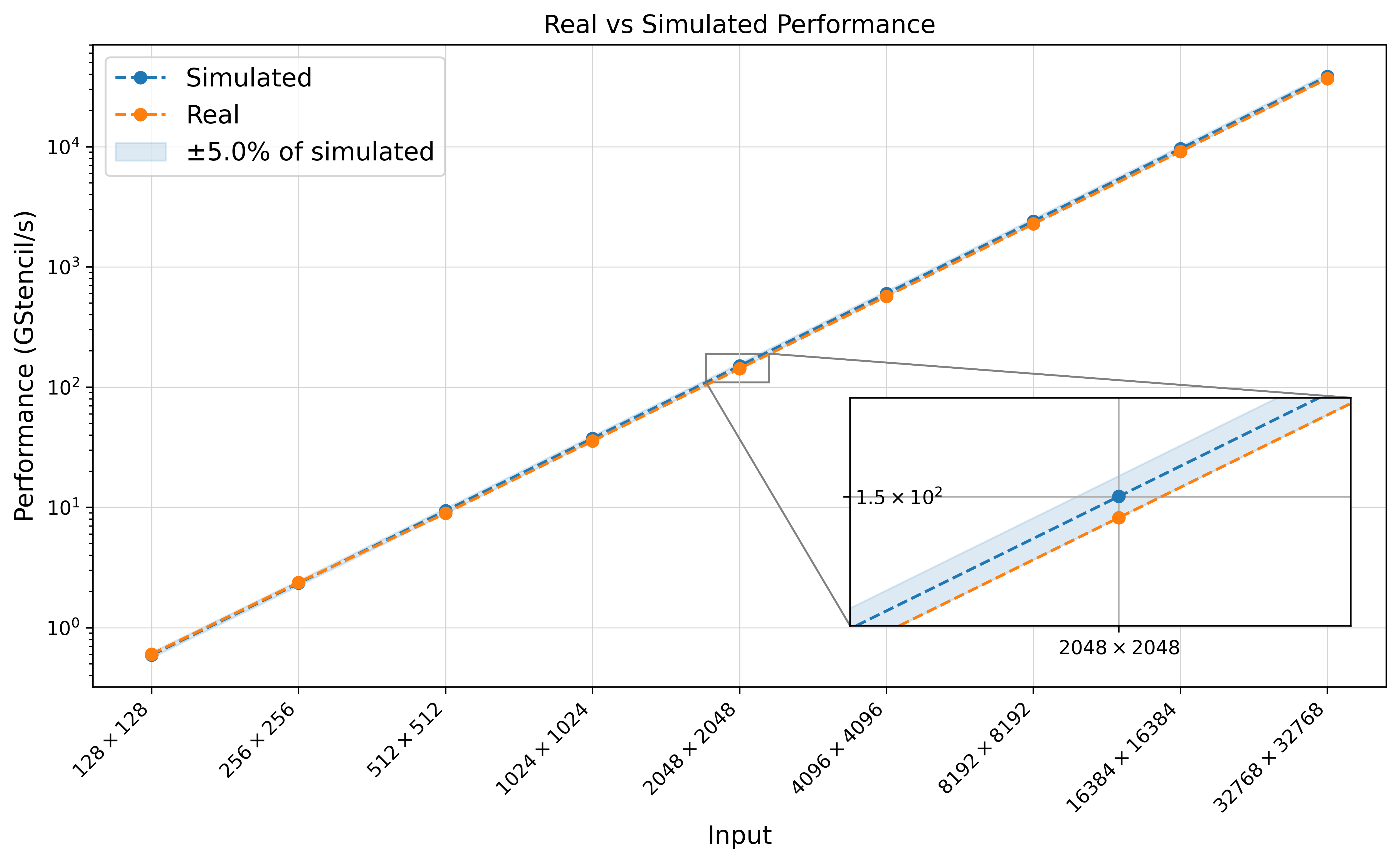}
        \caption{Validation of the simulator accuracy using the Star2d-1r pattern. The shaded band represents the 95\% confidence interval of deviations between real and simulated performance.}
        \label{fig:sim_vs_real}
    \end{figure}

Evaluating the Wafer-Scale Engine employs a hybrid approach combining direct hardware execution with cycle-accurate simulation. To ensure reliability, the simulator's accuracy was validated using the Star2d-1r pattern, with results showing that nearly all real measurements fall within a 5.02\% confidence interval of the simulated predictions (Fig.~\ref{fig:sim_vs_real}).

This high degree of correlation confirms the simulator as a reliable proxy for estimating the performance of other stencil patterns across the tested range. The validation spanned from a $2\times2$ PE grid up to a $512\times512$ configuration, corresponding to 262,144 active processing elements.

\begin{figure}[htbp]
        \centering
        \includegraphics[width=0.9\linewidth]{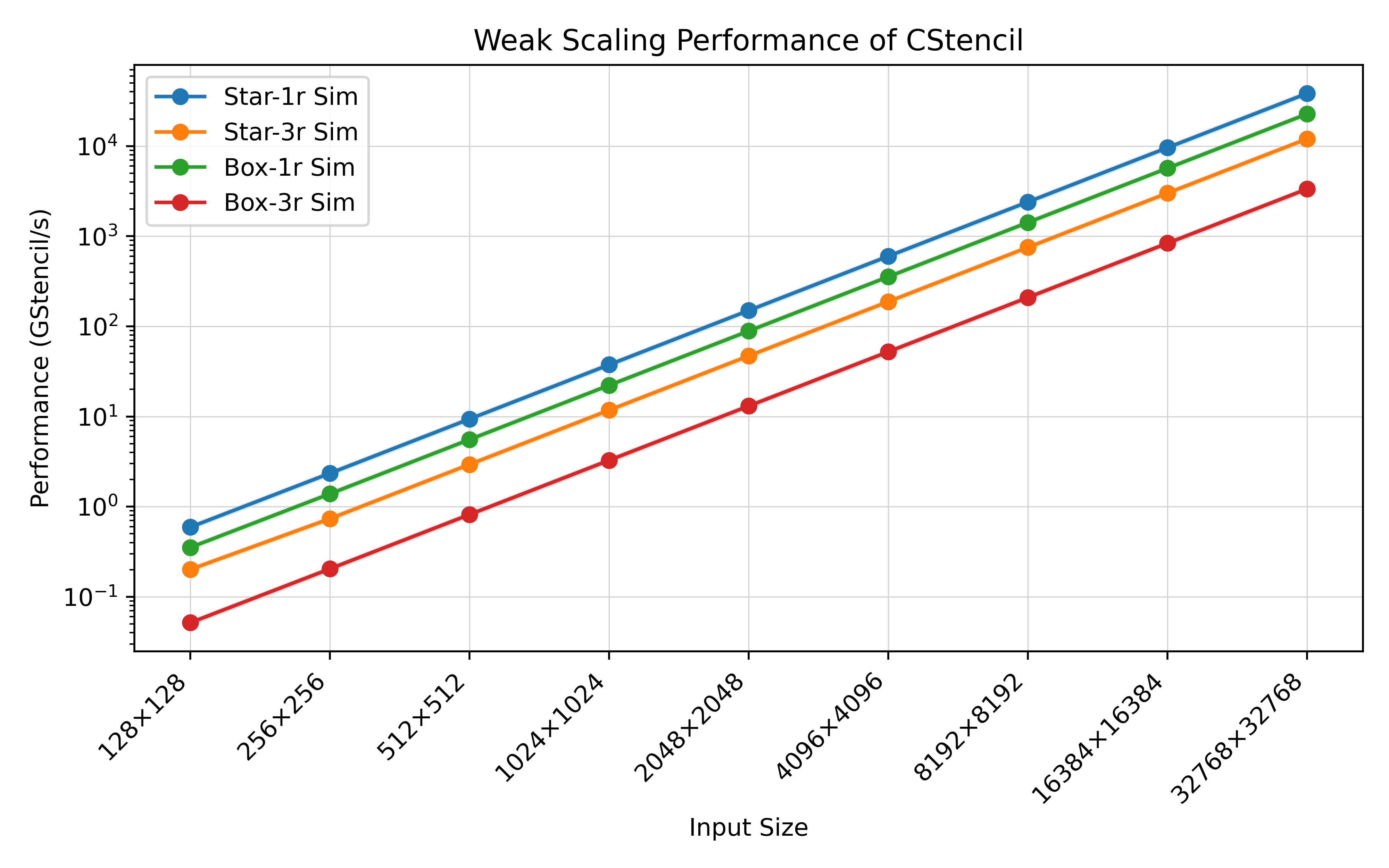}
        \caption{Weak scaling of various stencil patterns on WSE-3.}
        \label{fig:wse_sim_scaling}
    \end{figure}

Using the validated simulator, the weak scaling analysis was extended across multiple stencil patterns on the WSE-3 architecture. Each processing element (PE) was assigned a $64\times 64$ subgrid to maximize the utilization of the 48\,KB local SRAM. 
The results in Figure~\ref{fig:wse_sim_scaling} reveal near-perfect weak scaling across the tested domain sizes. This linear performance growth is attributed to the WSE-3’s massive on-chip memory bandwidth and an efficient mapping strategy that minimizes inter-PE communication overhead. The consistent scaling across different stencil complexities demonstrates the platform's general suitability for large-scale stencil workloads.

\subsection{Cross-platform Comparison: WSE-3 vs A100}
\begin{figure}[H]
    \centering
    \includegraphics[width=0.9\linewidth]{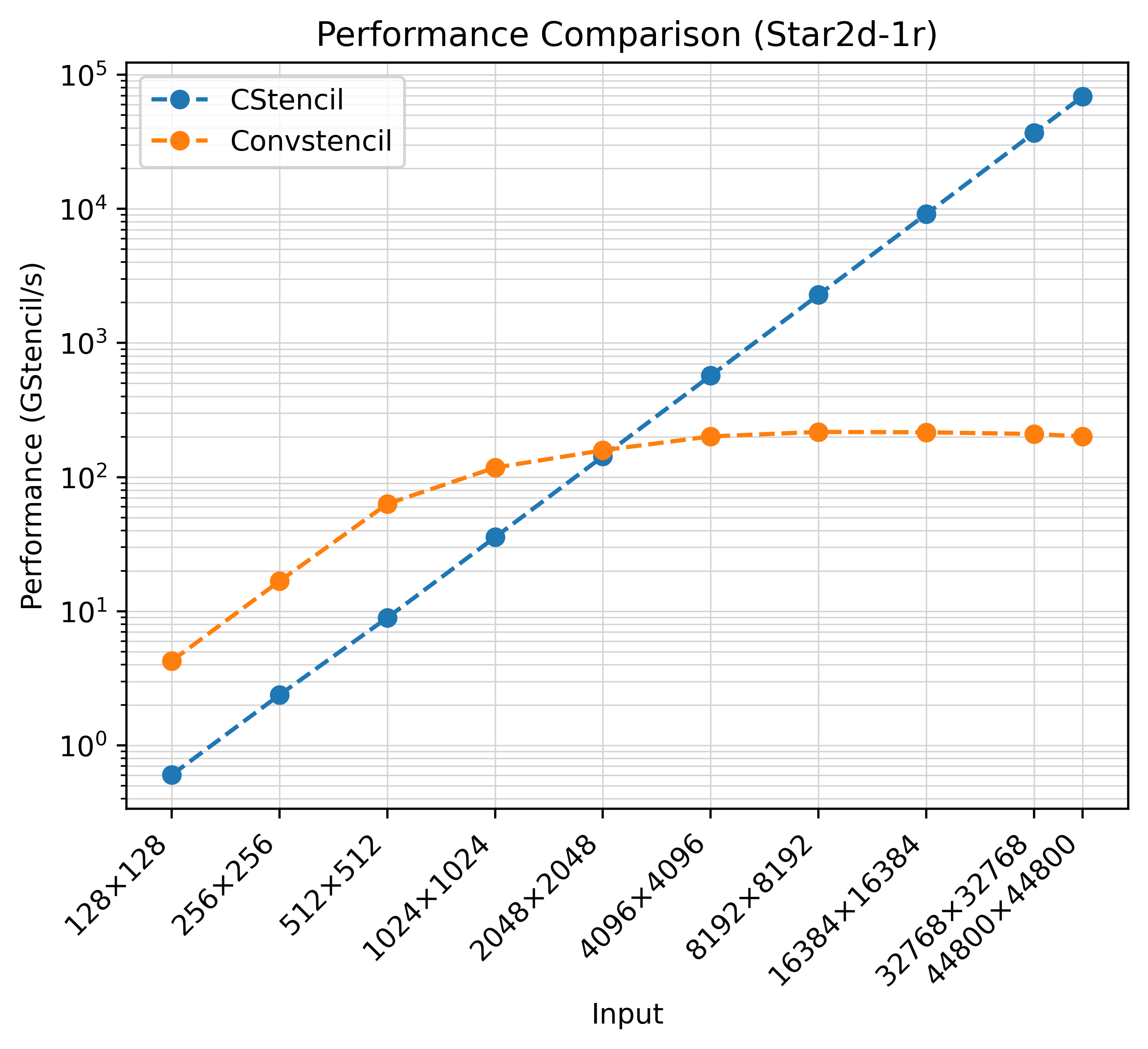}
    \caption{Absolute performance of CStencil (WSE-3) and ConvStencil (A100) on the Star2d-1r pattern.}
    \label{fig:max_input_timing}
\end{figure}

The absolute performance comparison reveals a significant advantage for the WSE-3, which achieves a 342$\times$ speedup over the A100 GPU at the largest tested scale for the Star2d-1r pattern (Fig.~\ref{fig:max_input_timing}). Even assuming an optimistic 8$\times$ throughput increase for a theoretical FP32 ConvStencil GPU implementation, the WSE-3 would still be characterized by 40$\times$ higher performance. This gap is primarily driven by the WSE-3's high on-chip SRAM bandwidth, overcoming the HBM bottlenecks that constrain the GPU's ConvStencil implementation. 

While the GPU performs better at grid sizes below $2048\times2048$ due to underutilization of the WSE-3's processing elements, the wafer-scale architecture's near-linear scaling allows it to quickly overtake the GPU as more PEs are involved.

\begin{figure}[h]
    \centering
    \includegraphics[width=0.9\linewidth]{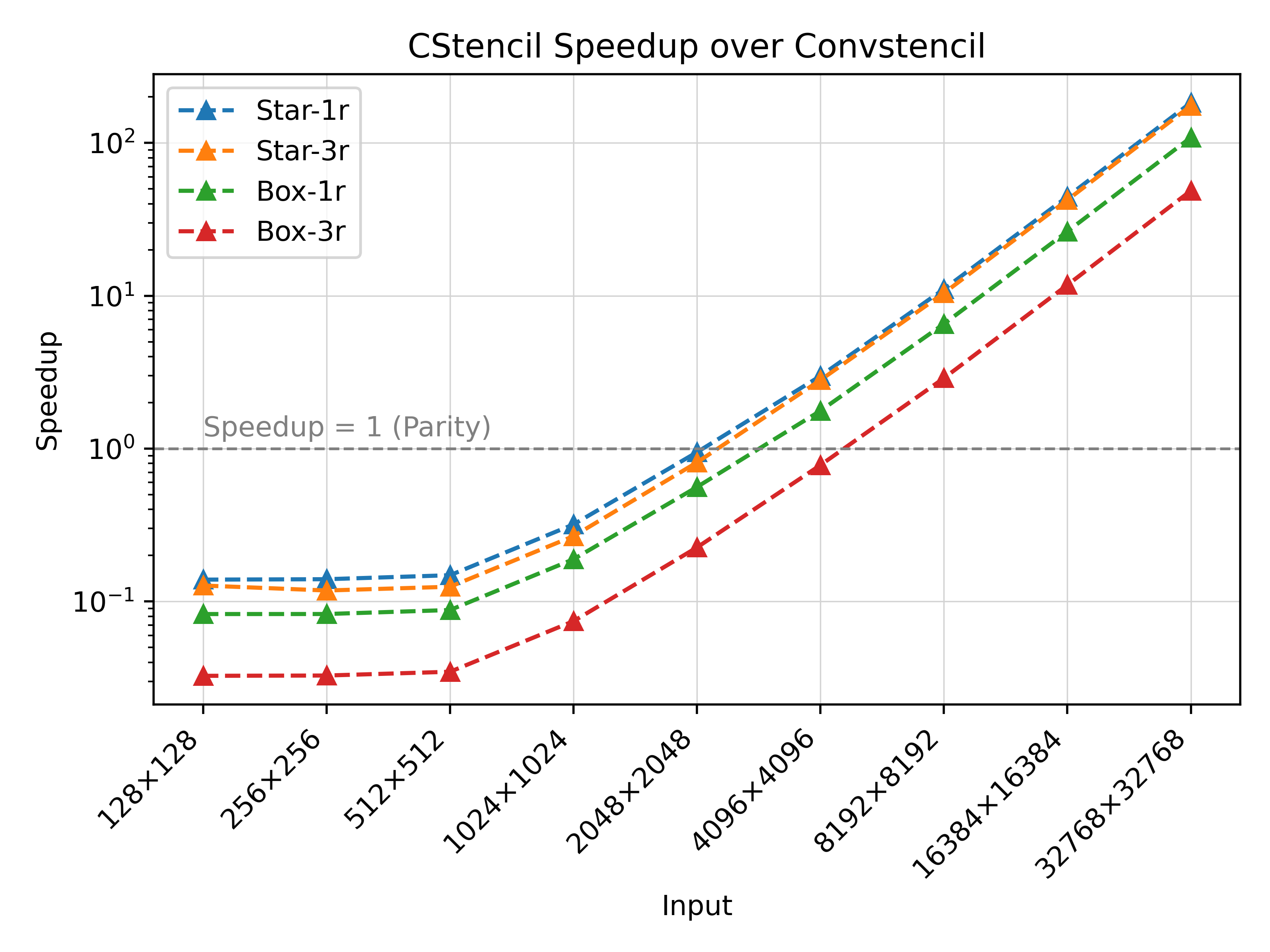}
    \caption{Speedup of CStencil over ConvStencil as a function of grid size for Star2d ($r=1,3$) and Box2d ($r=1,3$) patterns. The dashed line at $1.0\times$ represents performance parity.}
    \label{fig:speedup_curve}
\end{figure}

Figure \ref{fig:speedup_curve} shows the speedup of Cstencil (on WSE-3) over ConvStencil (on A100) for different stencil patterns and sizes. The analysis shows that even for high-intensity patterns like Box2d-3r, which are optimized for Tensor Core execution, the GPU implementation is eventually outperformed by the aggregate compute power and localized memory access of the WSE-3. Ultimately, the Star2d-1r pattern exhibits the highest speedup; because the implementation is compute-bound on the WSE-3, and the lightest computational workload allows the architecture to maximize its throughput relative to the memory-constrained GPU.

\subsection{Roofline Analysis}
To illustrate the fundamental bottlenecks of each architecture, we performed a roofline analysis using the Star2d-1r pattern on a $44,800\times44,800$ grid.
The arithmetic intensity (AI) for CStencil is calculated based on its 9 FLOPs per update and 10 memory accesses (5 reads, 5 writes) to SRAM, yielding $AI_{CStencil}\simeq0.23$\,FLOP/Byte. For the $700\times700$ PE subarray used in this study, the theoretical peak performance at 875.0\,MHz is 0.858 PFLOP/s.

\begin{figure}[h]
    \centering
    \includegraphics[width=0.9\linewidth]{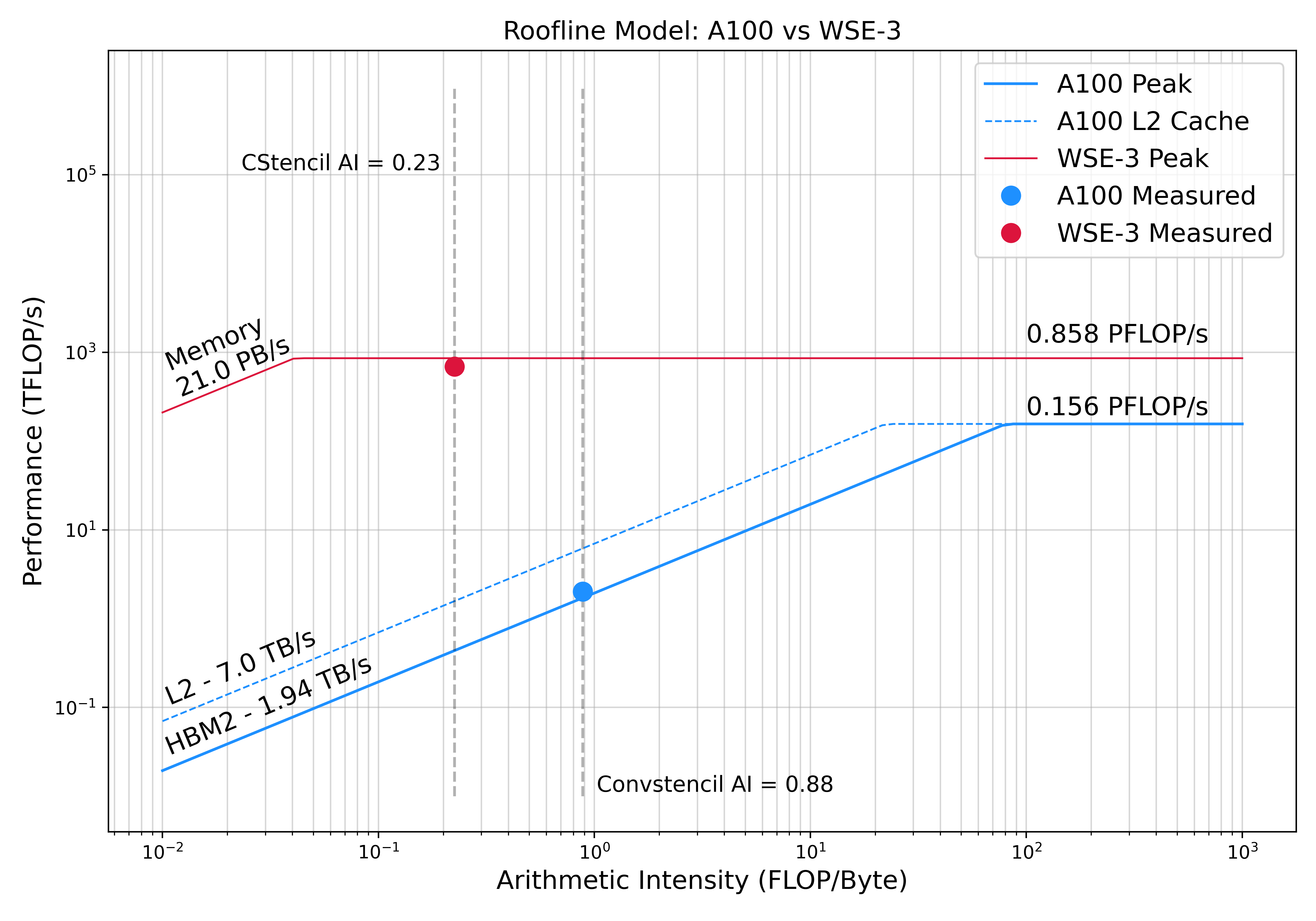}
    \caption{Roofline comparison of A100 and WSE-3 for the Star2d-1r stencil on a $44{,}800 \times 44{,}800$ grid, highlighting differences in memory and compute bottlenecks.}
    \label{fig:roofline}
\end{figure}

The comparison in Figure \ref{fig:roofline} highlights the architectural shift: ConvStencil on the A100 is strictly memory-bound by HBM bandwidth. In contrast, CStencil on the WSE-3 leverages massive on-chip SRAM bandwidth to operate significantly closer to its theoretical compute peak, demonstrating that wafer-scale integration effectively eliminates the traditional memory wall for stencil computations.

\section{Discussion and Conclusion}
\label{sec:conclusion}

The implementation of \textit{CStencil} on the Cerebras WSE-3 demonstrates a fundamental shift from memory-centric to communication-centric computation, addressing the "Memory Wall" that limits stencil kernels on conventional architectures. By leveraging a large on-chip distributed SRAM, the WSE-3 allows the entire computational domain to reside physically adjacent to the Processing Elements (PEs). This eliminates the performance bottleneck associated with transferring data between global memory (HBM) and computing units, a constraint that dictates the utilization rates of traditional GPUs.

Our analysis reveals that while GPUs remain superior for smaller domains due to mature SIMT scheduling and lower latency overhead, the WSE-3 exhibits nearly linear weak scaling. Because stencil computations rely on local, nearest-neighbor communication, the 2D mesh interconnect enables the WSE-3 to maintain a constant execution time per iteration as the problem size and PE count increase, resulting in a speedup of up to 342$\times$ on large domains. This architectural alignment makes wafer-scale integration particularly effective for extreme-scale simulations that would otherwise be throttled by PCIe or NVLink bottlenecks in multi-GPU configurations.

However, these gains come with significant trade-offs in programming complexity and I/O management. Developing for the WSE-3 requires low-level manual data routing and explicit synchronization via the Cerebras Software Language (CSL), which lacks the abstraction, tooling maturity, and portability of established ecosystems like CUDA. This places a higher cognitive burden on the programmer and increases development time, particularly for applications with irregular communication patterns. Future work to mitigate these challenges includes the development of higher-level abstractions and programming models capable of mapping general-purpose workloads onto efficient wafer-scale dataflow executions.

Ultimately, this work confirms that AI-oriented hardware can be effectively repurposed for high-performance scientific computing. By adapting grid-based algorithms to a dataflow paradigm, the WSE-3 establishes new attainable performance levels for memory-bound scientific applications. While it does not replace the GPU for general-purpose tasks, it offers a structurally sound approach for large-scale simulations in fields such as fluid dynamics and computational physics, where massive on-chip throughput is paramount.


\bibliographystyle{IEEEtran}
\bibliography{main}

\appendices

\section{MMA Operations on Tensor Cores}
\label{app:mma_operations}

The design of high-performance kernels utilizing hardware accelerators requires a deep integration with Matrix Multiply-Accumulate (MMA) units. On NVIDIA architectures, these operations are performed by Tensor Cores using \textit{fragments}, which are small, fixed-size blocks of a matrix loaded into registers specifically for fast matrix arithmetic.

When referring to an MMA operation, the fundamental computation is expressed as:
\[
\mathbf{D}_{m\times n} = \mathbf{A}_{m\times k} \times \mathbf{B}_{k\times n} + \mathbf{C}_{m\times n},
\]
where $\mathbf{A}$ and $\mathbf{B}$ are input fragments, $\mathbf{C}$ is the accumulation fragment, and $\mathbf{D}$ is the output fragment. $\mathbf{C}$ and $\mathbf{D}$ can refer to the same fragment in memory.

The dimensions ($m, k, n$) of the MMA fragments are determined by their data precision. Table~\ref{tab:wmma_fragments} outlines the dimensions for the precisions relevant to this study. In practice, only limited combinations of multiplicand and accumulator types are supported due to architectural hardware constraints.

\begin{table}[ht]
\centering
\caption{MMA fragment dimensions for different precisions.}
\label{tab:wmma_fragments}
\begin{tabular}{c|ccc}
\toprule
Precision & $m$ & $k$ & $n$ \\
\midrule
\texttt{double} & 8 & 4 & 8 \\ 
\texttt{float}  & 16 & 8 & 16 \\ 
\texttt{tf32}   & 16 & 8 & 16 \\
\bottomrule
\end{tabular}
\end{table}

Not all GPU architectures support every fragment type. For instance, the NVIDIA A100 lacks native support for pure single-precision (FP32) MMA operations for multiplicands. Consequently, implementations must fall back to mixed-precision, utilizing the TensorFloat (\texttt{tf32}) type for multiplicands. \texttt{tf32} is a single-precision format optimized for Tensor Cores that employs a reduced 10-bit mantissa to balance computational throughput and numerical accuracy.

In CUDA, MMA operations are issued at the warp level via the WMMA API \cite{cuda-wmma}. All WMMA instructions are warp-level collectives, meaning the entire warp participates in each instruction. When loading or storing data, each thread in the warp is responsible for a specific portion of the fragment. Since the exact mapping between threads and data elements is handled by the hardware, the programmer operates on fragments as a whole. Consequently, it is not possible to load or store a portion of data smaller than the defined fragment size.

\end{document}